\documentstyle[12pt]{article}

\newcommand{\be}{\begin{equation}}
\newcommand{\ee}{\end{equation}}
\newcommand{\bea}{\begin{eqnarray}}
\newcommand{\eea}{\end{eqnarray}}

\newcommand{\m} {{\tilde m}}
\newcommand{\V} {{\cal V}_n}
\newcommand{\T} {\tau}
\newcommand{\R} {\rho}

\begin{document}

\begin{center}
\begin{large}
{\bf  dS-Holographic $C$-Functions \\}
{\bf with a \\}
{\bf  Topological,  Dilatonic  Twist \\}
\end{large}  
\end{center}
\vspace*{0.50cm}
\begin{center}
{\sl by\\}
\vspace*{1.00cm}
{\bf A.J.M. Medved\\}
\vspace*{1.00cm}
{\sl
Department of Physics and Theoretical Physics Institute\\
University of Alberta\\
Edmonton, Canada T6G-2J1\\
{[e-mail: amedved@phys.ualberta.ca]}}\\
\end{center}
\bigskip\noindent
\begin{center}
\begin{large}
{\bf
ABSTRACT
}
\end{large}
\end{center}
\vspace*{0.50cm}
\par
\noindent
Recently,
the holographic aspects of asymptotically de Sitter
spacetimes have generated substantial literary interest. 
The plot continues in this paper, as we investigate 
a certain  class of dilatonically  deformed
 ``topological'' de Sitter solutions (which were introduced in
hep-th/0110234). Although such solutions possess a
detrimental cosmological
singularity,  their interpretation from a holographic perspective
remains somewhat unclear.  The current focus is on the
 associated  generalized $C$-functions,
which are shown to maintain their usual monotonicity properties
in spite of this  exotic framework. These findings suggest that
such  topological solutions may still play a role
in our understanding of quantum gravity with a positive cosmological
constant.

\newpage

\section{Introduction}
\par

There is considerable observational evidence that  the physical universe has
 a positive (albeit, disturbingly small) cosmological constant 
\cite{bops}.  This observation is  at least partially
responsible for the  recent flurry of  investigations
into asymptotically de Sitter  spacetimes.
In particular,  the various holographic aspects of de Sitter (dS) space
have garnered much attention. (See Ref.\cite{med2} for
a list of relevant citations. 
Also see 
Refs.\cite{bmsx,svx,caix,halx,myux,onx,dasx,bckzx,lmmx,nsx,lvl} for
more recent work.)
\par 
The focal  point of dS-based holography has been a conjectured
duality  between asymptotically dS  spacetimes and  conformal
field theories (CFTs) \cite{str}. This dS/CFT duality
can be viewed as an  analogy  to the celebrated anti-de Sitter 
(AdS)/CFT correspondence \cite{mal,gub,wit} (which, in turn,
is an explicit realization of the  renowned holographic
principle \cite{tho,sus}). 
The  CFT in a dS-inspired duality, just like  its AdS analogue,
 lives on an asymptotic boundary
of the bulk spacetime. However,  contrary to the AdS case,
the dS boundary is spacelike (located at temporal infinity) and the   
dual CFT is necessarily a Euclidean one. These distinctions can
be attributed to the absence of both a globally timelike Killing
vector and a spatial infinity in asymptotically dS spacetimes \cite{ssv}.
\par
The various  investigations into
 dS holography have, of course, lead to many interesting
deductions and observations.  At the forefront of these is
 Bousso's realization of an entropic upper bound \cite{bou}. 
More specifically, the entropy of pure
dS space serves as an upper bound on the total entropy
that can be stored in any spacetime with a positive cosmological 
constant.\footnote{Note that the validity of the Bousso bound does depend on
some form of the positive energy condition \cite{wald}.}
With guidance from the ``Bousso bound'', Balasubramanian, de Boer and Minic
\cite{bdm} have  proposed a similar upper limit 
on the total mass of an asymptotically dS spacetime.
In particular, these authors have conjectured that
any such spacetime whose conserved mass\footnote{The ``BdBM bound''
implies a specific definition for the   mass \cite{bdm,bk}. 
This definition
can be viewed as an appropriate  generalization of 
the  Brown-York  quasi-local energy \cite{by}.}
exceeds that of pure dS space will contain a naked cosmological singularity.
\par
For the sake of argument, let us  accept the conjectured  mass
bound  as being a  true property of asymptotically
 dS spacetimes (as  recent analysis does seem
to support \cite{cmz}). In this case,  from a bulk viewpoint,
the implications are quite severe; 
 a violation of this bound results in  a type of singularity that
is considered to be  {\it non grata} in  most 
cosmological models. On the other hand, from a holographic
perspective, the implications are somewhat murkier.
As pointed out by Ghezelbash and Mann \cite{gm}, for a hypothetical observer
located on an asymptotic spacelike boundary (i.e., a ``CFT observer''),
any such singularity will remain causally  hidden behind the 
cosmological horizon. That is to say,  any quantity  measured
by this observer depends only on the boundary theory; in fact,
the observer need not be aware that an interior region even  exists.
To reword this in a philosophical sense,  what exactly constitutes
the ``true physical'' picture:
the boundary theory or the bulk (or both or neither)? 
\par
Given the stated ambiguity, one might argue that bound-violating
asymptotically dS solutions should not be dismissed {\it a priori}.
Such solutions have been explicitly formulated in a paper
by Cai, Myung and Zhang \cite{cmz} (with their original motivation
being to test the mass-bound conjecture). One of these so-called
``topological'' de Sitter (TdS) solutions can effectively be obtained 
with a sign reversal (in the mass term) of a more conventional 
Schwarzschild-dS 
solution.  As a consequence, the black hole horizon
disappears, leaving behind a  naked singularity enclosed by
the usual cosmological horizon.\footnote{It should be kept in mind
that a TdS solution can have a spherical, hyperbolic or
flat (cosmological) horizon geometry. This is contrary to
the Schwarzschild-dS case, which only permits a spherical horizon geometry. 
It is this diversity in TdS horizon geometries that prompted
the topological nomenclature  in analogy with Ref.\cite{bir}.}   
Furthermore, it can readily be confirmed that, for any TdS solution,
 the conserved  mass  (in accordance with the definition of Ref.\cite{bdm})
does indeed exceed that of its purely dS counterpart. 
\par
Since the original presentation by Cai {\it et al}. \cite{cmz},
some subsequent papers have considered   the implications  
of a possible  TdS/CFT duality \cite{cai,med,myung,med2}.
For the most part, this duality would appear to be preferential
to its Schwarzschild-dS counterpart;\footnote{In Ref.\cite{myung}, however,
the author argues against TdS solutions in a dynamical-boundary 
scenario. We elaborate on this point in the final section.} 
inasmuch as  the CFT energy
can only be positive in the TdS case \cite{danxx,cai}. (Significantly, a 
negative energy implies a non-unitary theory.) 
In view of this desirable feature, we  argue that 
 TdS solutions  merit further investigation, and proceed on
this basis.
\par
Ultimately, one might hope that quantum gravity can be used
to deduce the validity (or invalidity) of a field theory
that is   holographically dual to  a TdS bulk.
 However, as it is well known, a definitive theory of
 quantum gravity remains currently out of reach.
Even the most promising candidate,  string theory,
fails to provide
a suitable  description of dS space   \cite{ssv}.
Nevertheless, we propose that  much can still be learned
by subjecting the (conjectured) TdS/CFT  correspondence
to various holographic ``consistency checks''.
\par
With the above proposal in mind, let us consider the 
intriguing phenomena of
 holographic renormalization group (RG) flows.
Significantly, RG flows   
are commonly believed to  play a prominent role
in any holographic bulk/boundary  duality. 
(For instance, see Refs.\cite{ag,fgpw,sak,dvv,str2,bdm}.)
More specifically, as any relevant parameter of  a bulk spacetime
evolves, a RG flow is expected to be induced 
in the  dually related boundary theory.
This picture follows from the so-called ultraviolet/infrared correspondence
\cite{sw,pp}, which implies that bulk evolution towards
the infrared translates into   boundary flow towards the ultraviolet
and {\it vice versa}.
\par
What is particularly pertinent to the holographic-RG picture
is the existence of a generalized $C$-function (in analogy with
RG flows in a two-dimensional CFT context \cite{zam}).
Moreover, this $C$-function should exhibit various
monotonicity properties that are reflective of the
underlying ultraviolet/infrared  duality.
In view of all this, an 
appropriate analysis of a  prospective 
$C$-function  should  serve as a suitable ``laboratory''
for testing the viability of any conjectured  bulk/boundary   correspondence.
Just such an analysis, in a TdS context, will  serve as the
focal point of the current paper.
\par
If one sets out to ``test''  a  prospective
 $C$-function,  it  should be significantly
more informative  when non-trivial matter fields are allowed in the 
bulk theory.
 (See, for instance, Ref.\cite{kpsz}.)
For this reason, we will generalize our considerations
 to a  certain class of 
solutions that  can be viewed as  dilatonic deformations
of a  TdS geometry  \cite{cmz}.  
This  new class can alternatively  be viewed as  domain wall
solutions having a (flat) cosmological horizon. In fact, these solutions
are essentially analytic continuations of domain wall-black hole
spacetimes  that effectively describe  a truncated theory
of gauged supergravity \cite{cs,bks,co,cz}.
\par
Before discussing the content of the paper,  let us consider
a pair of caveats. Firstly,  
it should be kept in mind that  the dilatonically deformed solutions 
 are,
in general, {\bf not} asymptotically de Sitter.
Nonetheless,   in the limit of a constant dilaton field, a TdS
solution (with flat horizon geometry) will always be obtained \cite{cmz}.
Secondly, it is worth emphasizing
that the presence of bulk matter (in this case, the dilaton)
will typically break the conformal symmetry of a dual boundary
theory. That is to say, the holographic duality now under consideration
 can be viewed  as a dS analogue \cite{tow} 
 of the domain wall/quantum field theory (QFT)
correspondence \cite{bks,bbhv,co}.\footnote{Notably, the 
domain wall/QFT duality (assuming its validity) 
includes  the AdS/CFT  correspondence
as a very special case.}

\par
The remainder of this paper is organized as follows.
In Section 2, we begin by introducing the relevant 
 action and 
formulating the ($n$+2-dimensional) bulk solutions
of interest:   dilatonic deformations 
of  a  topological de Sitter spacetime
with a flat cosmological horizon \cite{cmz}. 
(We will subsequently refer to these as ``DTdS'' solutions.)
In what is essentially a review of material in Ref.\cite{cmz}, we then go
on to calculate the  quasi-local stress tensor \cite{by} and conserved
mass of a DTdS  solution.  Notably,
this calculation necessitates that a surface counterterm 
be added to the action  for the purpose of regulating
infrared divergences \cite{hs,bk,ejm,bdm,gm}.
\par
In Section 3, we consider a (presumably) dual, Euclidean QFT
that lives on an asymptotic boundary of the bulk spacetime.
In particular, we obtain explicit expressions for the
 QFT stress tensor and thermodynamics by
making appropriate identifications with  properties
of the DTdS bulk.
A Cardy-like form \cite{car} for the QFT entropy is then verified.
This result can be viewed as a generalization of  
the Cardy-Verlinde formula \cite{ver,sv}, as appropriate
for a flat horizon geometry \cite{kpsz}.\footnote{The Cardy-Verlinde
formula has already  been generalized for a multitude of holographic
scenarios.  Consult  Refs.\cite{jingx,jing} for a list of relevant citations.}
Significantly  to later analysis, we also identify the Casimir
entropy \cite{ver} of the QFT.
\par
The focus of Section 4 is on prospective  (generalized) $C$-functions
 in a DTdS/QFT holographic framework. Here,
we consider a pair of prescriptions for the quantity of interest.
The first is based on a  formula that expresses the $C$-function
 in terms
of local bulk geometry. (See, for instance, Refs.\cite{fgpw,sak}.) 
The second follows from the premise that the Casimir entropy 
of a  boundary theory  \cite{ver} can be regarded, quite literally, as 
a Cardy-like ``central charge'' \cite{car} and, hence, $C$-function \cite{zam}.
Both of the prescribed forms are  rigorously tested to see
if they evolve  monotonically  with respect to variations
in relevant parameters.  
Given  a few justifiable assumptions,
we are able to demonstrate that this is, indeed, always the case.
\par
Finally, Section 5  provides a summary and further discussion;
including  a brief account on   dynamical-boundary
scenarios.

\section{Dilatonic Deformations of TdS Spacetime}

\par
In this section, we will begin by formulating the bulk theory of
interest; namely, a dilatonically deformed  
``topological-de Sitter'' solution with a flat horizon geometry \cite{cmz}.
Keep in mind that this ``DTdS'' solution   
describes a domain wall spacetime
with a cosmological (but no black hole) horizon.

\subsection{Domain Wall Action and Solutions}
 
\par
To start off,  let us consider an action 
that describes an $n$+2-dimensional dilaton-gravity theory
with a Liouville-like potential. More specifically:
\be
I={1\over 16\pi G}\int_{\cal M}d^{n+2}x\sqrt{-g}\left[R-{1\over 2}
\left(\partial \phi\right)^2+V_{o}e^{-a\phi_o}\right]+I_{GH},
\label{1}
\ee
where ${\cal M}$ represents the manifold,
 $G$ is the $n$+2-dimensional Newton constant, and where
$V_o$ and $a$ are to be regarded as positive constants.
Note that we have also included the Gibbons-Hawking surface
term, $I_{GH}$, which is necessary for a well-defined 
variational principle on the boundary  of the manifold \cite{gh}.
This surface term takes the form:
\be
I_{GH}={1\over 8\pi G}
\int_{\partial{\cal M}}d^{n+1}x\sqrt{\left|h\right|}{\cal K},
\label{2}
\ee
where ${\cal K}$ is the trace of the extrinsic curvature on the boundary
($\partial {\cal M}$).
\par
Interestingly, the above action (\ref{1})  is known to  effectively describe
a truncated theory of gauged supergravity \cite{cs,bks}. Given this
pedigree, $V_o$ and $a$ can be directly expressed
in terms of ${\cal N}$ and $p$, where ${\cal N}$ is
the number of D$p$-branes in the originating theory \cite{co}.
\par
A certain  class of domain wall-black hole solutions (with Ricci-flat
horizons)  
has been found
for this action \cite{cz}.  
The associated metric and dilaton  can be expressed 
by way of the following formalism:
\be
ds^2=-f(r)dt^2+f^{-1}(r)dr^2+R(r)^2dx^2_{n},
\label{3}
\ee
\be
f(r)={V_oe^{-a\phi_o}r^{2N}\over nN\left[N(n+2)-1\right]l^{2N-2}}-
{mr^{1-nN}\over \sqrt{2nN(1-N)}},
\label{4}
\ee
\be
R(r)={r^N\over l^{N-2}},
\label{5}
\ee
\be
\phi(r)=\phi_o+\sqrt{2nN(1-N)}\ln(r).
\label{6}
\ee
\par
In the above,
 $\phi_o$ and $m$ are non-negative constants
of integration, 
 $dx^2_n$ is the dimensionless line element of
an $n$-dimensional Ricci-flat spacetime, and
 $l$ is some
length parameter  (insuring correct dimensionality)
that  will subsequently be set to unity.
(Note that $m=0$ corresponds to  a purely domain wall spacetime.)
 Furthermore, the parameter
$N$  has been defined in accordance with:
\be
a={\sqrt{2nN(1-n)}\over nN}.
\label{7}
\ee
In view of  Eq.(\ref{4}), the following bound should
be imposed 
 on $N$:
\be
(n+2)^{-1}\leq N\leq 1.
\label{7.5}
\ee
\par
For later convenience, let us  rewrite Eq.(\ref{4}) in the following form: 
\be
f(r)=br^{2N}-{\m\over r^{nN-1}},
\label{8}
\ee
where:
\be
b\equiv {V_oe^{-a\phi_o}\over nN\left[N(n+2)-1\right]},
\label{9}
\ee
\be
\m\equiv {m\over\sqrt{2nN(1-N)}}.
\label{10}
\ee
 It is now quite evident that the special case of
$N=1$ describes, with a suitable renormalization of $m$,
  an $n$+2-dimensional Schwarzschild-AdS black hole.
\par
When $N<1$,  this class of solutions is no longer asymptotically AdS.
However, one can still obtain a well-defined
quasi-local  stress 
tensor  \cite{by},
provided that an appropriate surface  counterterm has first been added to
the action \cite{hs,bk,ejm,co}. 
The conserved
mass ($M$) can be directly calculated via this stress-energy tensor, 
and one finds \cite{cz}:
\be 
M={nN\V\m \over 16\pi G}, 
\label{11}
\ee
where $\V$ is the  volume of $dx^2_n$ (i.e., the dimensionless
volume of the domain wall). Note that there is no 
 vacuum ($m=0$) contribution to the mass by virtue of the 
locally flat solution space 
\cite{ejm}.
\par
Now let us suppose that $V_o< 0$. In this case, the action (\ref{1}) can
still be  effectively viewed as the truncation of a  gauged
supergravity theory \cite{tow}.  Let us
further assume that $m$ (or ${\tilde m}$) $\leq 0$  and then,  
 for sake of convenience,
  redefine $V_o\rightarrow -V_o$ and
 ${\tilde m}\rightarrow -{\tilde m}$.   The prior solution
 remains unchanged except for the metric function $f(r)$,
which  should now  be revised as follows:
\be
f(r)={\m\over r^{nN-1}}-br^{2N}.
\label{12}
\ee
\par
For the special case of $N=1$, the revised solution   describes
a topological de Sitter spacetime \cite{cmz} with a flat (cosmological)
 horizon geometry.
For general $N$ (but  satisfying Eq.(\ref{7.5})),
this  dilatonically deformed solution is  no 
longer asymptotically dS, but it  
does, in fact, still  possess   a cosmological horizon.  Denoting the horizon
 location by
 $r=r_c$, we have:
\be
r_c=\left[{\m\over b}\right]^{1\over N(2+n)-1}.
\label{13}
\ee
It should be kept in mind that, for any allowed $N$, the cosmological horizon
encloses a naked singularity.
\par
The associated thermodynamics of any cosmological horizon
can be obtained via standard identifications \cite{gh2}. For
a DTdS solution in particular, 
 the horizon temperature and entropy are  respectively
calculated as follows:
\be
T_H=-{1\over 4\pi}\left.{df\over dr}\right|_{r=r_c}={b\left[N(n+2)-1
\right]\over 4\pi} r_c^{2N-1},
\label{14}
\ee
\be
S_H={''area''\over 4G}={\V r_c^{nN}\over 4 G}.
\label{15}
\ee

\subsection{Quasi-Local Stress Tensor and Mass}

\par
With the above thermodynamic identities, 
it can readily be shown that the first law of horizon thermodynamics,
$dM=T_HdS_H$, is uniquely satisfied (up to the usual constant)
 for  $M$ as given by Eq.(\ref{11}). 
However, it is still instructive to  derive
  $M$ on a more fundamental level; namely, as  
 the conserved charge associated with time translation.
Note that the following analysis  essentially  reviews 
a calculation that is found  in Ref.\cite{cmz}.
\par
Let us begin here by considering  the spacetime
 outside of the cosmological
horizon (i.e., $r>r_c$) and, thus, 
appropriately shielded from the naked singularity.
  (Ultimately, we are interested
in the asymptotic limit of $r\rightarrow\infty$;
that is, future spacelike infinity or ${\cal I}_+$ \cite{ssv}.)
 The coordinates $r$ and $t$ change their character
in crossing over the horizon (from spacelike to timelike and
{\it vice versa}), and so, for illustrative purposes, 
we will relabel these as  
$r\rightarrow\T$ and $t\rightarrow\R$.  Given that $\T\geq \T_c=r_c$, 
the DTdS metric (\ref{3},\ref{12})  takes on the following suggestive form:
\be
ds^2=-f^{-1}(\T)d\T^2+f(\T)d\R^2 +R(\T)^2dx^2_{n},
\label{16}
\ee
\be
f(\T)= b\T^{2N}-{\m\over \T^{nN-1}} \quad \geq 0.
\label{17}
\ee
\par
Being somewhat more specific,  we now
 focus
  on an
$n$+1-dimensional spacelike boundary ($\partial{\cal M}$). The
boundary geometry can suitably be described 
 by the following metric:\footnote{Note the following conventions.
 Greek indices
will imply  coordinates of the $n$+2-dimensional manifold, 
``low-order'' Roman indices
($a$, $b$, ...) will imply  boundary coordinates, and higher-order
Roman indices ($i$, $j$, ...) will imply coordinates on the
$dx^2_n$ hypersurface.}
\be
ds^2_{B}= h_{ab}dx^a dx^b = f(\T)d\R^2+ R(\T)^2dx^2_{n},
\label{18}
\ee
where, for any specific boundary,
 $\T$ is  fixed at some  value  greater than $\T_c$.
\par
For a calculation of conserved charges (via a generalized 
 Brown-York
treatment \cite{by}), it is necessary to add an appropriate surface counterterm
to the action in question \cite{bk,ejm,bdm,gm}. (The premise being that
the counterterm can be used to cancel off the infrared divergences of
a given gravitational action \cite{hs}.)
Generally speaking,
 such a counterterm
is a necessarily  complicated expression involving the intrinsic curvature
of the boundary metric. However, for the current investigation, this intrinsic
curvature vanishes (by virtue of $dx^2_n$ being Ricci flat)
and the counterterm Lagrangian should  reduce into a relatively
simple, single-term form.
In analogy with the  deformed AdS-Schwarzschild  theory \cite{co}, 
a suitably defined counterterm
has already  been identified  for the DTdS form of the action  \cite{cmz}.
This result can be expressed as follows:
\be
I_{ct}=-{1\over 8\pi G}\int_{\partial {\cal M}} d^n xd\R \sqrt{h}
{n\over L},
\label{19}
\ee
where we have defined:
\be
L^{-1}\equiv N \sqrt{b} \T^{N-1}.
\label{20}
\ee  
Here, $L$ can be viewed  as the DTdS  generalization of
a  dS curvature radius.
 Although  $L$ generally  varies
throughout the  manifold  due to the factor of 
$\T^{N-1}$,\footnote{Actually, the
factor of $\T^{N-1}$ was omitted in the original
presentation \cite{cmz}. Such a factor is, however, necessary so as to ensure
the correct dimensionality of the counterterm Lagrangian. 
Other possibilities, such as  $l^{N-1}$ (where $l$ is the length
parameter that has been set to unity), do not achieve the desired
cancellation of infrared divergences.}
 the analogy with dS space
 becomes evident when  $N=1$.  In this limiting case, 
 $I_{ct}$  clearly reduces
to  the  anticipated asymptotically dS  form \cite{bdm}.
\par
One can calculate
the (so-called) quasi-local stress tensor \cite{by}  by
varying 
 the total action, $I+I_{ct}$ (including the Gibbons-Hawking
surface term \cite{gh}),   with respect to the boundary metric \cite{bk,bdm}.
Applying this prescription, we find:
\be
T_{ab}=-{2\over \sqrt{h}}{\delta (I+I_{ct})\over \delta h^{ab}}
= {1\over 8\pi G}\left[{\cal K}_{ab}-{\cal K} h_{ab}+{nNb^{1\over 2}\over
\T^{N-1}}h_{ab}\right].
\label{21}
\ee
\par
To compute the extrinsic curvature (and its trace),  we
will employ 
a  standard definition:
${\cal K}_{ab}= h^{\mu}_{a}h^{\nu}_{b}\nabla_{\mu}\eta_{\nu}$,
where $\eta^{\mu}$ is the  unit normal vector to the boundary.
With  the normal suitably taken  to be 
 $\eta^{\mu}=f^{1\over 2} \delta^{\mu}_{\T}$, 
some straightforward calculation yields:
\be
{\cal K}_{ij}= \delta_{ij}N f^{1\over 2}\T^{2N-1},
\label{22}
\ee
\be
{\cal K}_{\R\R}={1\over 2}f^{\prime}f^{1\over 2},
\label{23}
\ee
\be
{\cal K}= {\cal K}^{a}_{a}
={nNf^{1\over 2}\over \T}+{1\over 2}{f^{\prime}\over f^{1\over 2}}.
\label{24}
\ee
Note that  a prime indicates differentiation with respect to $\T$.
\par
Substituting the above results  into 
the stress tensor (\ref{21}) and
considering the large $\tau$ limit, we eventually obtain
the following  leading-order relations:
\be
T_{ij}= -\delta_{ij}{(2N-1)\m \T^{-(n-1)N}\over 16\pi G b^{1\over 2}},
\label{25}
\ee
\be
T_{\R\R}= { nN\m b^{1\over 2}\T^{-(n-1)N}\over 16\pi G}.
\label{26}
\ee
\par
Let us now focus our attentions on  the calculation of the conserved mass. 
Once again, we will  utilize
the techniques of Refs.\cite{bk,bdm}, which constitute  a generalization
of the Brown-York methodology \cite{by}.
\par
First of all, let us  discuss
the  general formalism for an
arbitrary ($n$+2-dimensional) spacetime. It is appropriate to
 consider an $n$+1-dimensional
 boundary ($\partial{\cal M}$)  and an $n$-dimensional  spacelike
surface, $\Sigma$, that is enclosed within $\partial{\cal M}$.
Ideally, given a Killing vector ($k^{a}$) 
 that generates an isometry of the boundary geometry,
one would like to calculate the associated conserved charge ($Q$).
This task can be accomplished with the following formula \cite{bk}:
\be
Q=\oint_{\Sigma} d^n\phi \sqrt{\sigma} u^{a} k^{b}T_{ab},
\label{27}
\ee 
where
the metric on $\Sigma$
has been parametrized according to
 $ds^2_{\Sigma}=\sigma_{ij}\phi^i\phi^j$, $u^{a}$ is
the unit normal vector  to $\Sigma$, and
$T_{ab}$ is the quasi-local stress tensor associated
with  $\partial{\cal M}$.  
\par
Next,  let us specialize  to 
an asymptotically dS spacetime and a calculation of the conserved mass
(i.e., $Q=M$). 
Normally, one associates the conserved
 mass with a globally  timelike Killing vector.
Although
there is no such Killing vector for an asymptotically dS spacetime,
it has  satisfactorily been  shown that one can consider the analytic
continuation of $\partial_t$ \cite{bdm}; where
$t$ is the temporal coordinate inside of the cosmological
horizon.  If we take $\R$  as being the analytic continuation
of $t$ (outside of the cosmological horizon) and conveniently
choose  $u^a$ to be proportional to the relevant Killing vector,
then Eq.(\ref{27}) takes on the following form:
\be
M=\oint_{\Sigma} d^n\phi \sqrt{\sigma} N_{\R}u^{\R}u^{\R} T_{\R\R},
\label{28}
\ee 
where the Killing vector has been normalized according to
$k^{a}=N_{\R}u^a$. ($N_{\R}$ can be identified with the
usual ``lapse function'' of the boundary metric.)
\par
Finally, we are in a position to consider the model currently
under investigation.
Assuming that the above formalism can directly be adapted into 
a DTdS  framework, \footnote{It is worth repeating
that  DTdS solutions are not 
 asymptotically dS, except in the special case of $N=1$.} 
we can accordingly re-express Eq.(\ref{28})   as follows: 
\be
M=\V\T^{nN}f^{-{1\over 2}}(\T)T_{\R\R}.
\label{29}
\ee
To obtain this form,  the following identities have been applied:
$\oint_{\Sigma} d^n\phi\sqrt{\sigma}=\V\T^{nN}$, 
$u^a=\delta^{a}_{\R} f^{-1/2}$, and $N_{\R}=f^{1/2}$.
\par
Substituting Eq.(\ref{26}) into the above relation and taking the 
asymptotic limit, 
we ultimately  find:
\be
\lim_{\T\rightarrow\infty} M={nN\over 16\pi G}\V\m.
\label{30}
\ee
Fortunately, this outcome  agrees with  the priorly quoted result (\ref{11}) 
and, thus, with the first law of DTdS horizon thermodynamics.

\section{Euclidean QFTs on the Boundary}

It has been suggested that the AdS/CFT correspondence \cite{mal,gub,wit}
is really  just a special case of a more general
holographic duality; namely,  
the domain wall/QFT correspondence
\cite{bks,bbhv,co}. By way of analogy, one might  argue that the dS/CFT
correspondence \cite{str}  is also a special case
of a more encompassing duality.  That is to say,
there  may  exist
 a  dual relationship
between certain domain wall spacetimes and Euclidean  QFTs \cite{tow,cmz}. 
Although somewhat speculative, we
will adopt  this  viewpoint for the duration of the paper.

\subsection{QFT Geometry and Stress Tensor}
 
\par
With the above discussion in mind,  let us now  consider 
a   Euclidean QFT that  lives on an asymptotic 
boundary of a  DTdS bulk spacetime.   Presumably (or perhaps naively),
the bulk and boundary theories should be  dually related 
in a holographic sense.
Keep in mind that, for the very special case of $N=1$, the (TdS) bulk
theory is asymptotically dS, the
 boundary theory is a conformal one, and a holographic duality
appears to be in evidence   \cite{cai,med,myung,med2}.
\par
As is typically the case in holographic bulk/boundary dualities, the
metric of the QFT in question will be fixed,
up to a conformal factor,
as the metric on an asymptotic boundary of the bulk spacetime.
By analogy with   Ref.\cite{sv}, we invoke:
\bea
ds^2_{QFT}&=&\gamma_{ab}dx^adx^b \nonumber\\
&=& \lim_{\T\rightarrow\infty} {1\over \T^{2N} b} ds^2 \nonumber \\
&=& d\R^2 +b^{-1} dx^2_n,
\label{31}
\eea
where $ds^2$ (in the second line) is the metric defined by
 Eqs.(\ref{16},\ref{17}).  Let us re-emphasize that
 all current/future considerations are restricted 
to  the  region outside of the cosmological horizon (i.e., $\T\geq \T_c$).
\par
We can calculate the stress tensor (${\cal T}_{ab}$) of
the QFT by  way of  the following relation  \cite{mye}:
\be
\sqrt{\gamma}\gamma^{ab}{\cal T}_{bc}=\lim_{\T\rightarrow\infty}
\sqrt{h}h^{ab}T_{bc},
\label{32}
\ee
where $T_{ab}$ is the quasi-local stress tensor of
Eqs.(\ref{25},\ref{26}) and
the boundary metric, $h_{ab}$,  is  defined by Eq.(\ref{18}).
\par
Utilizing the above relation, we are able to deduce:
\be
{\cal T}_{ij}= -\delta_{ij}{(2N-1)\m b^{n-2\over 2}\over 16\pi G} ,
\label{33}
\ee
\be
{\cal T}_{\R\R}= { nN\m b^{n\over 2}\over 16\pi G}.
\label{34}
\ee
\par
It is interesting to note that the trace of the stress tensor,
${\cal T}={\cal T}^{a}_{a}$, only vanishes  in the special
 case of $N=1$. Reassuringly,  this special 
case describes an asymptotically dS
spacetime, and  so  the corresponding QFT should, indeed, be  a conformal
theory.  Of further interest,  the above results
imply the following equation of state for
the  QFT (with $\epsilon$ and $p$ respectively denoting
energy density and pressure):
\bea
\omega\equiv {p\over \epsilon} &=& -b{{\cal T}_{ij}\delta^{i}_{j}
\over {\cal T}_{\R\R}} \nonumber\\
&=& {2N-1\over nN},
\label{35}
\eea
where we have assumed  a perfect-fluid description.
As expected,
this equation  reduces to $\omega=1/n$ (i.e., radiative matter) when
the $N=1$ conformal theory  is realized.

\subsection{QFT Thermodynamics}

\par
Let us now consider the  thermodynamic properties of  this QFT 
and subsequently determine if they can  accommodate  a Cardy-Verlinde-like
 entropic form \cite{car,ver}.
As is the usual practice, we first rescale the QFT metric
so that the associated  boundary is located at a fixed  ``radial distance'',
{$\cal R$},  from the origin.\footnote{Strictly speaking,
${\cal R}$ represents temporal evolution when outside
of the cosmological horizon.}
This necessitates the following
conformal transformation: 
\be
ds^2_{QFT} \rightarrow b{\cal R}^2 ds^2_{QFT} = 
b{\cal R}^2 d\R^2 +{\cal R}^2 dx^2_n.
\label{36}
\ee
However, we have not yet realized  the desired form; that is:
\be
ds^2_{QFT} = 
 d\R^2 +{\cal R}^2 dx^2_n.
\label{37}
\ee
Evidently, 
 the Euclidean time coordinate
should be further  rescaled such that $\R\rightarrow b^{1/2}{\cal R}\R$.
It follows that, from a QFT perspective,
 the bulk  energy and temperature (having
units of inverse time)
 should be ``red shifted'' by a factor of
$\Delta=\left[b^{1/2}{\cal R}\right]^{-1}$.
\par
In accordance with the above discussion, the QFT thermodynamics
can be identified as follows:\footnote{Note that the 
entropy is universally unaffected by any such coordinate
rescaling \cite{wit}.}
\bea
E_{QFT}=\Delta M&=&{nN\V\m\over 16\pi G b^{1\over 2}{\cal R}}
\nonumber \\
&=& {nN\V b^{1\over2}\over 16\pi G {\cal R}} \T_{c}^{N(2+n)-1},
\label{38}
\eea
\be
T_{QFT}= \Delta T_H = {b^{1\over 2}\left[N(n+2)-1
\right]\over 4\pi {\cal R}} \T_c^{2N-1},
\label{39}
\ee
\be
S_{QFT}=S_H={\V \over 4 G} \T_c^{nN},
\label{40}
\ee
where we have incorporated Eqs.(\ref{13},\ref{14},\ref{15},\ref{30}).
It can  readily be  verified that these relations
satisfy the first law of QFT thermodynamics; that is, 
$dE_{QFT}=T_{QFT}dS_{QFT}$.  We also take note of  $E_{QFT}\geq 0$;
  which is  indicative of a  topological-dS, rather than
Schwarzschild-dS, holographic framework \cite{danxx,cai}. 
\par
In general and regardless of dimensionality,  the entropy of any
horizon and, by duality,  the entropy of its
 holographic  boundary theory
should be  expressible  in a  Cardy-like form 
\cite{carlip,solly}. That is to say, one might expect \cite{car}:
\be
S_{QFT}={2\pi\over n}\sqrt{{c\over 6}\left[L_o -{c\over 24}\right]},
\label{41}
\ee
where $L_{o}={\cal R}E_{QFT}$
and the ``central charge'',  $c$,  is directly proportional
to the Casimir (i.e., sub-extensive) energy of the boundary theory \cite{ver}.
However, given a bulk theory with a flat horizon geometry (which
implies a vanishing Casimir energy), the
associated QFT  can be expected to conform with the following
version \cite{kpsz,jing}:
\be
S_{QFT}={2\pi\over n}\sqrt{{c L_o\over 6}}.
\label{42}
\ee
As a consequence of this form,  $c$ is, in principle, proportional  to
an appropriately defined Casimir entropy \cite{ver} 
(which, unlike the Casimir energy,
remains   finite  and positive, regardless of the
horizon geometry \cite{kpsz,youm}).
\par
We find that the above thermodynamic relations do indeed
satisfy Eq.(\ref{42}) (with $L_{o}={\cal R}E_{QFT}$), as long as:
\be
c={3n\V\over 2 \pi G N b^{1\over 2}} \T_c^{N(n-2)+1}.
\label{43}
\ee
\par
The expected relation between the  Casimir entropy ($S_C$)
and the generalized  central charge is $S_C=\pi c/6n$ 
(for instance, \cite{kpsz}).  On this basis:
\be
S_C={\V\over 4  G N b^{1\over 2}} \T_c^{N(n-2)+1}.
\label{44}
\ee
It is hard to confirm the validity of this result,
insofar as  we are unable to calculate the Casimir entropy
by  more  direct means. (This is contrary to the usual scenario for
a spherical horizon:
 the Casimir entropy is directly proportional 
to  the  Casimir
energy, which represents a violation in the Euler identity \cite{ver}.)
Nonetheless, it is quite reassuring that  $S_C$ (as defined above)
reduces to its anticipated form in the special $N=1$ case \cite{med}.

\section{Generalized C-Functions}

With inspiration from the ultraviolet/infrared correspondence \cite{sw,pp},
it is commonly believed that
 evolution of  a bulk spacetime  will  give rise to
some form of  RG flow in its  holographically related boundary theory
(for instance, \cite{ag,fgpw,sak,dvv,str2,bdm}).
Moreover, there should exist some generalized 
 $C$-function (in analogy with  two-dimensional RG flows
\cite{zam}) that exhibits appropriate monotonicity properties
as the state of the system varies.
On the basis of such arguments,
the existence (or lack thereof) of a suitable $C$-function
should   serve as an appropriate litmus test for
 a conjectured bulk/boundary duality.
This philosophy, in a DTdS/QFT context,  will serve as
the  premise for the analysis that follows.  
\par
Given a bulk spacetime and its dually related boundary theory,
there are two commonly used prescriptions for the
generalized $C$-function.
We will examine both of these in turn.

\subsection{Bulk-Geometrical Prescription}

\par
Firstly, let us consider any bulk ($n$+2-dimensional) spacetime for which 
the metric  can
be expressed in the following domain wall-like form:
\be
ds^2=-dz^2 +e^{2A(z)}\left[dy^2+dx_{n}^2\right]. 
\label{45}
\ee
In this case, one expects  the existence of a generalized $C$-function
that is based on local bulk geometry and can be represented as follows 
(for instance, \cite{fgpw,sak}):\footnote{A very recent paper \cite{lmmx}
considered
a revised form  for this  $C$-function, which apparently has a 
wider range of applicability.
 However,
in the case of a flat horizon geometry (as is relevant to the current study),
 this  newer formulation
reduces to  Eq.(\ref{46}).}
\be
C\sim {1\over G \left[A^{\prime}(z)\right]^n}.
\label{46}
\ee
Note that  a prime now indicates differentiation with respect
to $z$.
\par
 The DTdS  bulk metric of Eq.(\ref{16})
can be cast into the above template (\ref{45}) by way
of  the following
identifications:
\be
dz= {1\over \sqrt{f(\T)}} d\T,
\label{47}
\ee
\be
y= {\sqrt{f(\T)}\over R(\T)}\R,
\label{48}
\ee
\be
A(z)=A(\T)=\ln[R(\T)]=N\ln(\T).
\label{49}
\ee  
\par
As it stands, such  a calculation of $C$ would not be
particularly enlightening. Nevertheless, we can still viably proceed
by first  assuming that the effective mass parameter, $\m$, is
much smaller than  the other relevant  scales.
(However, $\m$ should remain a  non-vanishing quantity,  so that a 
cosmological horizon is still in existence.)
It is significant that, with this assumption,  $f(\T)\approx b\T^{2N}$
becomes a valid approximation (cf. Eq.(\ref{17})).
\par
Applying the pertinent  approximation to  Eq.(\ref{47}),
we obtain the following useful relation:
\be
dz \approx {\T^{-N}\over b^{1\over 2}}d\T.
\label{50}
\ee
\par
In terms of the above formalism,
the   prescribed $C$-function (\ref{46}) now yields: 
\be
C\sim {\T^{n(1-N)}\over GN^n b^{n\over 2}}.
\label{52}
\ee
It is immediately  clear that
 $C$  increases monotonically  with increasing ``radial distance''
$\T$ (recalling that $N\leq 1$). Hence,  we have 
confirmed the anticipated ultraviolet/infrared duality \cite{sw,pp}. That
is, the infrared (large $\T$) limit of the bulk theory 
corresponds to the ultraviolet (large $C$) limit of the QFT 
and {\it vice versa}.  
\par
Interestingly, we see that $C$ becomes a constant (with
respect to variations in $\T$) when $N=1$. That is,
the constant-dilaton (TdS) theory
translates to a conformal fixed point of the holographic RG flow.
Clearly, this $N=1$ fixed point is an infrared one.
\par
It should also be  instructive  to  examine
the  behavior of the $C$-function  under variations in $N$.
Significantly, changes in $N$ reflect variations 
in the matter content of the theory. To put it another way,
as $N$ monotonically  decreases below its conformal value of $1$,  the
bulk scalar fields are effectively being ``turned on''
(cf. Eq.(\ref{6})). In fact, simple  analysis 
tells us that
 the dilaton field will continue to  ``grow''
until  $N=1/2$ has been reached. Thus, one might
expect $N=1/2$ to represent a fixed point in the associated RG
flow.
(Clearly, the conformal value, $N=1$,  serves
as the other fixed point.) We will provide further  support
for this claim  in the latter part of this section.
\par
To proceed along the suggested line, it is necessary
to re-express Eq.(\ref{52}) so that all $N$ dependence
is explicit.
We can accomplish this task by
substituting for $b=b(N)$  via Eq.(\ref{9}).  This process yields:
\be
C\sim {\T^{n(1-N)}\over G}\left[{nN(n+2)-n\over NV_o}\right]^{n\over 2},
\label{53}
\ee
where we have set $\phi_o=0$ for  sake of convenience.
\par
Given that  $C$ is  strictly a positive quantity, it is just as informative
(and substantially easier) to
consider variations in $\ln(C)$. Hence, it is useful to write:
\be
\ln(C) = n(1-N)\ln(\T) +{n\over 2}\ln[N(n+2)-1]
-{n\over 2}\ln(N) + constant.
\label{54}
\ee 
Varying this expression  with respect to $N$, we have:
\be
{\partial \ln(C)\over \partial N}= -\ln(\T)+
{n\over 2N}\left[N(n+2)-1\right]^{-1}.
\label{55}
\ee
\par
The above result indicates that, for sufficiently
large values of $\T$, $C$ is a monotonically decreasing 
function of $N$. Furthermore, since $1$ is an upper bound
on $N$,
 the conformal theory 
can be identified   as the infrared 
 fixed point with respect to  
  bulk-matter evolution.  Given that
the $N=1$ limit corresponds to an essentially matter-free
 theory, this identification
seems pleasantly intuitive. At this juncture, however,
the ultraviolet fixed point seems somewhat less clear.
\par
Before proceeding on to the next phase of the analysis,
let us comment on the condition of  ``sufficiently large
$\T$''. This constraint can be viewed as a  manifestation
of  a certain aspect of the  DTdS/QFT framework. 
In particular,  any external observer will  be unable to access
 information from  behind the cosmological
horizon and  the duality must, therefore,  naturally break down
when    $\T\leq\T_c$. To put it another way, $\T_c$ can
be viewed as a necessary  ultraviolet cutoff for the DTdS bulk 
or (by duality \cite{sw,pp}) an infrared cutoff for the QFT.

\subsection{Casimir-Entropic Prescription}
 
\par
Alternatively, given a holographic boundary theory,
  the Casimir entropy ($S_C$)
has also been  interpreted as a generalized $C$-function. (With 
considerable success; see, for
instance, Refs.\cite{kpsz,hal}).
This interpretation of $S_C$ follows directly from  its role as an effective
central charge \cite{car,zam} in the Cardy-Verlinde formula \cite{ver}. 
Recalling  Section 3, we have already identified the Casimir entropy
(\ref{44}) for
the  QFT of interest.  On this basis, let us now consider: 
\be
C=S_C={\V\over 4  G N b^{1\over 2}} \T_c^{N(n-2)+1}.
\label{56}
\ee
\par
Considering our ``gameplan'',
 this  $C$-function can most conveniently be  
expressed as an explicit function of
  $T_{QFT}$
and $N$. Applying Eq.(\ref{9}) for $b=b(N)$ and
Eq.(\ref{39}) for  the temperature, we  eventually obtain the following
expression:
\be
C= {\V\over 4G} \left[{n\over V_o}\right]^{nN\over 2(2N-1)}
\left[{N(n+2)-1\over N}\right]^{N(4-n)-2\over 2(2N-1)}
\left[4\pi {\cal R}T\right]^{N(n-2)+1\over 2N-1}.
\label{57}
\ee
Note that  $\phi_o$ has again been set to vanish  and  the
subscript ``QFT''  has been dropped  from the temperature.
\par
As discussed before, it is both convenient and sufficient to
consider the logarithm of $C$. 
Up to some irrelevant constant terms,  we find the following:
\bea
\ln(C) = -{nN\over 2(2N-1)} \ln\left({V_o\over n}\right)
&+&{N(4-n)-2\over 2(2N-1)}\ln\left[{N(n+2)-1\over N}\right]
\nonumber \\
&+&{N(n-2)+1\over 2N-1}\ln(4\pi{\cal R} T).
\label{58}
\eea
\par
Let us first consider varying $\ln(C)$ with respect to
the boundary radius (${\cal R}$):
\be
{\cal R}{\partial \ln(C)\over\partial {\cal R}}=
{N(n-2)+1\over 2N-1}.
\label{59}
\ee
If we assume that  (i) $n \geq 2-(1/N)$ and (ii) $N\geq {1\over 2}$,  
then the following bound can be established:
\be
{\partial \ln(C)\over\partial {\cal R}} \geq 0.
\label{60}
\ee
That is, the ultraviolet/infrared connection has (once again) been  verified.
Some commentary on the assumed conditions is, however, still in order.
\par
Condition (i) simply limits considerations  to
a bulk  theory of dimensionality  three or greater,
which obviously covers all  physically relevant dimensionalities.   
\par
Condition (ii) is  more difficult to interpret, given
that, strictly speaking, $N$ can take on values as low as
$(n+2)^{-1}$ (cf. Eq.(\ref{7.5})). Nonetheless, it is also
of relevance  that any value of $N$ below ${1\over 2}$
translates into a QFT with a  negative pressure; cf. Eq.(\ref{35}).
Interestingly, it has been argued that the 
domain wall/QFT correspondence
will break down  for these types of  negative-pressure states  \cite{co,cz}.
(This argument follows from an observation that gravity
fails to  decouple from the QFT  when the pressure falls below zero.)
In view of this consideration,  $N\geq 1/2$ seems 
to be  quite a natural constraint.  Furthermore,
this lower bound on $N$ supports a  prior hypothesis; namely, 
 that $N=1/2$ is
the most suitable candidate for an  ultraviolet fixed 
point. To reiterate,
this hypothesis is based on the observance that,
as $N$ decreases below $1$, the dilaton
will grow  until $N=1/2$  has been reached.  
\par
Next, let us consider how the  $C$-function evolves when the
 temperature is varied. 
It is clear that $C$  has the same functional dependence
on $T$ as it has on ${\cal R}$. Hence, imposing the
same justifiable  constraints as before, we  have:
\be
{\partial \ln(C)\over\partial T} \geq 0.
\label{62}
\ee
That is,  the C-function evolves monotonically with respect
to temperature and, moreover, 
 the QFT flows  towards
the ultraviolet as temperature increases. This outcome agrees with
the usual expectation  that   thermal excitations will induce
additional degrees of freedom.
\par
Finally, let us consider variations in the $C$-function with
respect to the parameter $N$; keeping in mind that a decreasing
 $N$ translates into  excitations of     the bulk scalar. 
By way of Eq.(\ref{58}), the following is found:
\bea
{\partial \ln(C)\over\partial N} 
= &-&{n\over (2N-1)^2}\left[ 
\ln\left(4 \pi {\cal R} T\right)
-{1\over 2}\ln\left({N(n+2)-1\over n N V_o^{-1}}\right)\right]
\nonumber \\
  &-&  
 {N(n-4)+2\over 2N(2N-1)}\left[N(n+2)-1\right]^{-1}.
\label{63}
\eea
It is evident that, for  both ``sufficiently large'' values of temperature
and $N\geq {1\over2}$, the following relation will always be satisfied:
\be
{\partial \ln(C)\over\partial[-N]} \geq 0.
\label{64}
\ee 
Hence, under suitable conditions,
  $C$  is a monotonically increasing function as
$N$ decreases. 
This result agrees with our previous finding; thus  reconfirming 
the intuitive  notion
of  bulk matter fields inducing a flow to the ultraviolet. 
\par
Let us now  comment on the most recently imposed conditions.
  $N\geq {1\over2}$ is just the previously discussed
 positive-pressure constraint, whereas 
 the  condition of large temperature can be 
justified  by virtue of  the following argument.
The Cardy-Verlinde formula, upon which this definition
of  $C$ has been based,   only
has validity in a  regime of large temperature \cite{ver}. 
In fact,  a breakdown can be expected in the Cardy-Verlinde formalism  
when  
${\cal R}T>>1$ is no longer satisfied \cite{ktxx,linxx,obmxx}. 
It is clear that   the same  limitation   can be  deduced
from  our findings.
\par
As an aside, it  would be interesting to determine
if there is some finite temperature at which 
Eq.(\ref{64}) does indeed begin to fail. Significantly, this  special
value of temperature
could be interpreted as the analogue of the Hawking-Page (Schwarzschild-AdS)
phase transition \cite{hpxx}. However, it appears that
 such a determination would require
a complicated  numerical analysis.  
\par
To briefly summarize, we have demonstrated that both definitions
 of the generalized $C$-function (\ref{46},\ref{56}) 
satisfy the  monotonicity properties
that would  be expected for  a bulk spacetime  with a QFT dual.

\section{Concluding Discussion}

In summary, we have been investigating into  the holographic properties
of  a special   class of domain wall solutions; these having
the distinction  of a  singularity enclosed by a cosmological
horizon.
Any of these bulk solutions can be interpreted as
 a dilatonic deformation of a topological de Sitter spacetime \cite{cmz}.
That is to say, in the limit of a constant dilaton field,
the once-deformed model will  describe an asymptotically de Sitter
solution with a cosmological singularity. As discussed in Section 1,
such a singularity may not be a particularly critical  
issue from the perspective
of an asymptotic boundary observer  \cite{gm}.
Also of note,   any of these deformed ``DTdS'' solutions
can be analytically continued into a domain wall spacetime 
 that effectively  describes a truncated 
theory of gauged supergravity \cite{cs,bks,co}.
\par
The initial phase of the analysis entailed strictly bulk considerations.
We began here by introducing  the relevant (arbitrary-dimensional) action,
which describes gravity coupled to a dilaton field with a Liouville-like
potential. For this  action, a certain class of domain wall-black hole 
solutions are known \cite{cz},  whereby  a trivial redefinition
of the potential  leads  to a DTdS solution space.
 With the bulk geometry precisely formulated, we 
went on to calculate the  quasi-local stress tensor   and
conserved  mass  \cite{by}
by way of the surface-counterterm method \cite{hs,bk,ejm,bdm}.
Significantly, the  calculated mass is precisely that which
 satisfies  the first law of cosmological horizon thermodynamics.
(We again note that this portion of the paper  was essentially
a review of  prior work \cite{cmz}.)
 \par
In the second phase of the analysis, we considered a Euclidean quantum
field theory that lives on an asymptotic boundary of the 
DTdS bulk spacetime. It was argued that the bulk and boundary theories
could well have a  dual  relationship \cite{tow} in analogy 
with the domain wall/QFT
correspondence \cite{bks,bbhv,co}. 
Utilizing  standard holographic relations \cite{mye,ver,sv},
we were able to identify the stress tensor and thermodynamic 
properties of this QFT. It was then shown that the QFT entropy 
satisfies a generalized form of the Cardy-Verlinde formula \cite{car,ver}.
Notably, this  generalization can be viewed as the appropriate one
for a flat horizon geometry \cite{kpsz,jing}.
On the basis of this formulation, we also identified the Casimir (i.e., 
sub-extensive) entropy of the QFT.
\par
The final phase of the analysis focused on generalized $C$-functions.
In this regard, we studied two  commonly used prescriptions:
(i) a formula that expresses $C$ in terms of local bulk geometry 
(for instance, \cite{fgpw,sak})
and (ii) an identification between $C$ and a  QFT-induced Casimir entropy
 (for instance, \cite{kpsz,hal}).
After formulating the appropriate expressions, we tested
these prescriptions by varying each of the $C$-functions
 with respect to  
relevant parameters.   For both versions, the  ultraviolet/infrared
correspondence \cite{sw,pp} was clearly established. (That is, $C$ increases
monotonically with respect to an increasing boundary radius.)     
Furthermore,  after imposing  a few  justifiable conditions, we
were able to show that $C$  increases monotonically
with respect to both increasing temperature and decreasing
 $N$; the latter being a parameter of the bulk theory.\footnote{For
sake of accuracy,  let us note that the temperature correspondence
was only verified  with the second prescription, whereas the $N$ 
correspondence was verified for both.}
 Both of these
outcomes  agree with prior expectations, given that
thermal excitations should  activate degrees of freedom
and a  decrease in  $N$ can be correlated  with the activation
 of bulk matter fields. 
\par
Also of interest,  a pair of   fixed points was identified
for the implied  renormalization group  flows.  In particular, the
conformal (or constant dilaton) case of $N=1$  
corresponds to an infrared  fixed point,
while $N=1/2$  describes   an  ultraviolet
fixed point. The   significance of the  ultraviolet limit is 
 that it describes  
a boundary theory with vanishing pressure; at which point,
a breakdown can be expected in 
the  (prospective)  DTdS/QFT duality \cite{co,cz}.
\par
In conclusion, the results of our analysis  are definitely in support 
 of a DTdS/QFT and  (hence) TdS/CFT  correspondence.
Further support for the latter duality
has come in prior studies; most notably,
it has been observed that a TdS bulk gives rise
to a positive-energy CFT \cite{cai,med2} (in direct contrast
to the  more conventional  Schwarzschild-dS bulk scenario \cite{danxx}). 
 However, we do not mean to imply that these studies
in any way verify the legitimacy of  topological de Sitter
solutions 
(i.e., asymptotically dS solutions  with a cosmological
singularity).
Rather, we view these positive outcomes as an argument that
such solutions should not be disregarded {\it a priori}.
Let us remind the reader  that an asymptotic boundary observer
would be causally inhibited from   accessing any information from behind
the TdS cosmological horizon; including  information about the
naked singularity \cite{gm}. Hence, the  potential legitimacy of
 TdS solutions seems
to depend on what constitutes the ``fundamental'' theory; the bulk or
the boundary.
It appears that this issue necessitates
further investigation.

\par
As a final consideration, we will provide  a somewhat brief account
 on  dynamical-brane  scenarios.\footnote{In this aside, we
prefer brane in lieu of boundary for no particular reason.}  
 In a prior paper \cite{med},  the implications
of a  brane  that evolves   
in a TdS background spacetime were considered.
More specifically, the methodology of Savonije and Verlinde \cite{sv},
for a radiation-dominated  brane universe  
moving in a Schwarzschild-AdS bulk, was suitably
generalized for a TdS framework.\footnote{Note that 
 the Savonije-Verlinde program was first generalized for
an asymptotically dS bulk by Ogushi \cite{oguxx}.}
Remarkably, the desirable
features of the Savonije-Verlinde treatment do indeed persist 
for a TdS bulk. (These  features include various holographic
entropy bounds \cite{ver}, the saturation of these bounds when
the brane crosses the  bulk horizon, and the coincidence
of the FRW equations\footnote{FRW implies  
 the  Friedman-Robertson-Walker cosmological  equations, which
typically  describe brane dynamics.} 
 with CFT thermodynamics at this saturation point.)
However,  it  has subsequently 
been noted by Myung \cite{myung}
that, for a TdS bulk, 
the  induced brane metric and (hence) Hubble parameter are
defined with respect to  Euclidean cosmological time.
When one continues back to Lorentzian cosmological time,  the FRW equations
are then  expressed in terms of an  undesirable negative energy density.
This problem could be  circumvented by adding suitable
matter to the brane; however, this is just the type of fine-tuning
mechanism that one usually tries to avoid.  
\par
If,  for  sake of argument,   the above criticism is disregarded,
then it is not difficult to further generalize the Savonije-Verlinde program
(and its pertinent outcomes) \cite{sv} to the case of a dilatonically deformed 
TdS bulk spacetime. One needs only to  follow the procedure outlined
by Cai and Zhang \cite{cz} for the analogous case of a  
``dilatonically deformed  Schwarzschild-AdS'' bulk spacetime
(i.e., the spacetime  described by 
 Eqs.(\ref{3}-\ref{6})). The only significant difference would be
in the definition of the cosmological time parameter ($\eta$).
More precisely, Eq.(3.4) from their paper:\footnote{We have
slightly modified the notation of Ref.\cite{cz} for
the purpose of (hopefully) avoiding confusion.}
\be
f(r)\left({dt\over
d\eta}\right)^2 -
{1\over f(r)}\left({dr\over d\eta}\right)^2=1 \quad\quad\quad
f(r)=br^{2N}-{\m\over r^{nN-1}},
\ee
 which leads to an induced brane metric of:
\be 
ds^2=-d\eta^2+{\cal R}^2(\eta)dx^2_n,
\ee
should be replaced with:
\be
f(\T)\left({d\R\over
d\eta}\right)^2 -
{1\over f(\T)}\left({d\T\over d\eta}\right)^2= 1 \quad\quad\quad
f(\T)= b\T^{2N}-{\m\over \T^{nN-1}},
\ee
which then leads to the following Euclidean form:
\be 
ds^2=d\eta^2+{\cal R}^2(\eta)dx^2_n.
\ee
(Note that $\T\geq\T_c$ is mandatory, whereas any $r\geq 0$ is allowed.) 
After making this substitution, one can essentially follow Ref.\cite{cz} 
in a straightforward manner. Their results (which, for the most part,
agree with those of Savonije and Verlinde \cite{sv})
will  persist unfettered for the DTdS 
case.\footnote{However, one should keep in mind
that, for the  DTdS case, the Hubble constant
is  Euclidean (rather than Lorentzian)   and the bulk horizon 
is a cosmological one (rather than a black hole).}
\par
Finally, let us comment on one of the more interesting outcomes
of Ref.\cite{cz}, which applies vicariously to the  deformed
TdS model. As it so happens, the resultant FRW equations imply
the following equation of state on the brane:
\be
\omega={p\over \epsilon}={1\over nN}.
\ee
This only agrees with our prior finding
({\it viz}. Eq.(\ref{35})):
\be
\omega = {2N-1\over nN}
\ee
in the special conformal case of $N=1$.
As argued by Cai and Zhang, this discrepancy can be attributed to
an {\it effective} gravitational constant 
that varies in time \cite{cz}:
\be
G_{n+1}={(n-1) N b \over\left[{\cal R}(\eta)\right]^{{1-N\over N}}}G,
\ee 
as well as the contributions from  a dynamical dilaton
field.\footnote{From the perspective of a dynamical-brane observer,
the dilaton field will naturally be viewed  as a function
of the cosmological time parameter.}
\par 
 The dynamical contributions from  gravity and the dilaton also seem
to  compensate for the effects of the apparently non-radiative matter 
(since, in general, $\omega
\neq n^{-1}$). That is to say,
  the 
holographic bounds and brane-horizon coincidences
observed by Savonije and Verlinde \cite{sv} 
can only be expected to persevere in a radiation-dominated  universe
 \cite{youmxx}; so it seems strange that, for the most
part, they do so here. We hope to investigate this matter
in a future work.

\section{Acknowledgments}
\par
The author  would like to thank  V.P.  Frolov  for helpful
conversations.



\end{document}